Photobiological effects at Earth's surface following a 50 pc Supernova

Brian C. Thomas

Washburn University, Department of Physics and Astronomy, Topeka, KS; 1700 SW College Ave., Topeka, KS 66604; 785-670-2144; brian.thomas@washburn.edu

Abstract:

We investigated the potential biological impacts at Earth's surface of stratospheric $O_3$ depletion caused by nearby supernovae known to have occurred about 2.5 and 8 million years ago at about 50 pc distance. New and previously published atmospheric chemistry modeling results were combined with radiative transfer modeling to determine changes in surface-level Solar irradiance and biological responses. We find that UVB irradiance is increased by a factor of 1.1 to 2.8, with large variation in latitude, and seasonally at high latitude regions. Changes in UVA and PAR (visible light) are much smaller. DNA damage (*in vitro*) is increased by factors similar to UVB, while other biological impacts (erythema, skin cancer, cataracts, marine phytoplankton photosynthesis inhibition, and plant damage) are increased by smaller amounts. We conclude that biological impacts due to increased UV irradiance in this SN case are not mass-extinction level, but might be expected to contribute to changes in species abundances; this result fits well with species turnover observed around the Pliocene-Pleistocene boundary.

1. Introduction

Recent work (Binns et al., 2016; Breitschwerdt et al., 2016; Fimiani et al., 2016; Fry et al., 2016; Ludwig et al., 2016; Wallner et al., 2016) has strongly supported earlier claims (Benitez et al., 2002; Knie et al., 1999, 2004) of at least one, and probably several, supernova explosions within 50-100 pc from Earth around 2.5 and 8 Ma. Thomas et al. (2016) and Melott et al. (2017) modeled a number of terrestrial effects of both photon and charged particle radiation from such supernovae. As noted by others before (Gehrels et al., 2003; Reid et al., 1978), depletion of stratospheric ozone is an important result of these events. Destruction of stratospheric $O_3$ allows greater penetration of Solar UVB radiation (280-315 nm), which is known to have damaging effects on life on Earth's surface and in the oceans.

In this work, we draw on results of Thomas et al. (2016) and Melott et al. (2017) and methods developed in Thomas et al. (2015) and Neale and Thomas (2016) to investigate changes in surface-level Solar irradiance and subsequent photobiological effects associated with $O_3$ depletion following a supernova at about 50 pc.

2. Methods

Here we use previously completed atmospheric chemistry modeling described in Thomas et al. (2016) and Melott et al. (2017). That modeling was completed using the Goddard Space Flight Center (GSFC) 2D atmospheric chemistry and dynamics model, with vertical profiles of SN-induced ionization calculated from modeled spectra of received radiation. We briefly describe the models here and refer readers to those papers for further details.

The GSFC atmosphere model has been described extensively elsewhere (see Ejzak et al., 2007; Thomas et al., 2007; Thomas et al., 2005; and references therein). The model's two spatial dimensions are altitude and latitude. The latitude range is divided into 18 equal bands and extends from pole to pole. The altitude range includes 58 evenly spaced logarithmic pressure levels (approximately 2 km spacing) from the ground to approximately 116 km. A lookup table is used for computation of photolytic source term, used in calculations of photodissociation rates of atmospheric constituents by sunlight. The model includes winds, small-scale mixing, solar cycle variations, and heterogeneous processes (including surface chemistry on polar stratospheric clouds and sulfate aerosols). The model contains 65 chemical species; a "family" approach is used for the transport of most species. Families include, for instance, $O_x$ [$O^3$, O, O($^1$D)], $NO_y$ (N, NO, $NO_2$, $NO_3$, $N_2O_5$, $HNO_3$, $HO_2NO_2$, $ClONO_2$, $BrONO_2$), $HO_x$ (H, OH, $HO_2$), $Cl_y$ (Cl, ClO, HCl, HOCl, $ClONO_2$), etc. Several other species are transported separately. We use the model in a pre-industrial state, with anthropogenic compounds (such as CFCs) set to zero.

Atmospheric ionization for use in the model is computed separately and then used as a source of $NO_y$ and $HO_x$ (both important families in ozone depletion reactions). Ionization by SN CRs in this work was computed using tables from Atri et al. (2010) which give ionization for primaries with energy from 300 MeV to 1 PeV, for 46 altitude bins, from the ground to 90 km. In this work, we used primary proton energies between 10 GeV and 1 PeV.

Supernovae are classified into a number of different types, depending on their spectra. In general, SNe produce radiation across the electromagnetic spectrum, with time-varying spectra that depend on the particular type. We chose type IIP as the most likely type associated with the events of interest (for more details see Fry et al., 2016 and Thomas et al., 2016). Thomas et al. (2016) and Melott et al. (2017) found, first of all, that the total energy in photons for the particular type of supernova most likely to have occurred at 2.5 Ma at 50-100 pc is too small to have a significant impact on atmospheric chemistry. On the other hand, supernovae also accelerate charged particles (mostly protons) which become what we observe as cosmic rays (CRs). Both high energy photons and CRs contribute to atmospheric ionization, leading to destruction of $O_3$ in the stratosphere. It is the cosmic ray flux that is most significant in the cases considered here. Thomas et al. (2016) and Melott et al. (2017) used CR spectra generated by simulating transport of protons from the SN through the interstellar medium to Earth; details on the methods and resulting spectra can be found in those works. Results in Thomas et al. (2016) for a 100 pc supernova indicate that $O_3$ depletion is not large enough to cause significant increases in Solar UVB, while Melott et al. (2017) found more significant depletion for a 50 pc supernova.

For the 50 pc supernova, two cases were considered, varying in the details of the magnetic field environment in the space between Earth and the supernova. The most realistic scenario is denoted "Case B" in Melott et al. (2017). Both the overall flux and energy spectrum of CRs received at Earth varies over time. Melott et al. (2017) examined atmospheric changes at 300 years after the first arrival of CRs from the Case B 50 pc supernova. This time was chosen since it gives the largest amount of ionization in the stratosphere, thereby maximizing the impact on $O_3$ column density. For this work we have performed additional atmospheric simulations using the 100 year and 1000 year ionization profiles from Melott et al. (2017).

In Melott et al. (2017) we made a very simple estimate of biological damage caused by increased Solar UVB using a "radiation amplification factor" (RAF; Madronich et al., 1998). The RAF method uses empirically determined values in a power-law scaling relationship to relate $O_3$ column density to biologically active UVB irradiance. We computed an increase of about 1.5 times for the 50 pc, Case B SN, at 300 years.

In this work we seek to improve on that rough estimate of biological impact. Thomas et al. (2015) found that for atmospheric chemistry changes associated with a gamma-ray burst (GRB), consideration *only* of changes in UVB under depleted $O_3$ does not give a full picture of the range of possible biological effects. While the RAF methodology is a good first-look at potential impacts, it neglects certain complexities. First, UVB is not the only biologically active irradiance band. Changes in the UVA (315-400 nm) and visible (400-700 nm; also referred to as "photosynthetically active radiation" or PAR) bands are also important for a number of biological impacts. For ionizing radiation events there is initial production of $NO_2$ which has a strong absorption band around 400 nm, affecting surface-level irradiance of UVA and PAR. In addition, $O_3$ has an absorption band around 600 nm (the "Chappuis band"). When signifcant $O_3$ depletion occurs, then, both UVB and PAR can be increased, while UVA may be reduced. These complexities can only be illuminated with full radiative transfer modeling and spectral biological weighting functions (BWFs).

Following Thomas et al. (2015), we use version 5.0 of the publically available Tropospheric Ultraviolet and Visible (TUV) radiative transfer model, downloaded from https://www2.acom.ucar.edu/modeling/tropospheric-ultraviolet-and-visible-tuv-radiation-model (Madronich and Flocke 1997). The model is 1D (vertical) and can be run at any specific location (in latitude and longitude) and time. It includes normal atmospheric constituents, as well as the option to specify different column densities for constituents such as $O_3$, $NO_2$ and $SO_2$. Aerosol and cloud parameters can be specified. We use the option to compute radiative transfer using the discrete ordinates 4-stream method. We have modified the code to simplify use with vertical constituent profiles generated by our atmospheric chemistry modeling, and to facilitate automation of runs over time and location. Aerosol and cloud parameter values are chosen to represent clear-sky "clean" conditions, as discussed in Thomas et al. (2015). All modeling presented here is done at local noon.

The TUV model produces spectral irradiance and integrated irradiances for UVB, UVA and PAR. In addition, many biological weighting functions are available within the model. A BWF quantifies the wavelength-dependent effectiveness for a biological outcome such as DNA damage. TUV convolves the computed spectral irradiance with built-in BWFs to produce weighted irradiances or relative impact. In this work, then, we improve on the RAF-based estimates of biological impact made by Melott et al. (2017), by using full radiative transfer modeling with latitude and time dependent atmospheric profiles of $O_3$ and $NO_2$, generating irradiance fields at Earth's surface and convolving those with BWFs to compute a variety of biological effects.

Some authors have claimed there is a correlation between cosmic ray flux and clouds in Earth's atmosphere (see, for instance, Marsh & Svensmark, 2000). However, others (Erlykin & Wolfendale, 2011; Sloan & Wolfendale, 2008) have questioned this correlation. While plausible physical mechanisms have been proposed, involving enhanced formation of aerosols and nucleation sites, (Svensmark et al., 2009; Tinsley, 2000), none has been experimentally established (Duplissy et al., 2010; Kirkby et al., 2011; Pierce & Adams, 2009; Wagner et al., 2001). While this possible effect on clouds and aerosols is relevant to our present study, with the large uncertainty surrounding the issue and no well-established physical mechanism, we could not accurately include such effects in atmospheric chemistry modeling. Furthermore, with no quantitative connection between ionization and aerosol/cloud characteristics, there is no way to realistically determine the appropriate parameter values for radiative transfer modeling. We therefore ignore any changes in aerosols or clouds proceeding from ionization in the atmosphere.

3. Results

Atmospheric simulations were run using ionization profiles for three time frames following arrival of SN CRs, as described above. The CR ionization was input as a steady-state source starting 1 year into the atmosphere model run. After about 10 years the atmosphere reaches a steady state with $O_3$ reduced compared to a control run without SN CR input. Since our runs are not connected to any particular dates, all time values shown in these figures are arbitrary and given simply to provide scale of the time frames involved.

Figure 1 shows the globally averaged change in $O_3$ column density, as a percent difference, after the atmosphere reaches an equilibrium state under a constant CR flux, for the 100 yr, 300 yr and 1 kyr cases. Time 0 is chosen arbitrarily during a steady-state model run for each case, after equilibrium is reached. The periodicity present is due to seasonality, since $O_3$ chemistry depends strongly on photolytic reactions.

Figure 2 shows the percent difference in $O_3$ column density as a function of latitude and time for the 300 year case (versus a control run). Here we see the approach to steady-state under a constant CR flux. Note that in Figures 1 and 2 the decrease in $O_3$ varies seasonally since concentrations are affected by photochemical reactions. For the 300 yr case $O_3$ depletion varies between about 15-40% over much of the globe. The latitude-time dependent change is similar for all three cases, with a maximum localized depletion of 38% for the 100 yr case, 43% for the 300 yr case, and 31% for the 1 kyr case.

Next, we present results of modeling changes in Solar irradiance at Earth's surface, and subsequent biological impacts. Given the results shown in Figure 1, we chose to focus on the 300 yr case, in the steady-state, with globally averaged $O_3$ depletion around 26% (Figure 1) and localized depletions > 10% for essentially the entire globe. These conditions persist for several hundred years, with gradual transition across the cases shown in Figure 1.

In Figure 3 we show the increase in UVB (as a ratio of the SN case to the control run) as a function of time, at latitudes between 5° and 55° North. We have chosen these latitudes as representative of the change; note in Figure 2 that the $O_3$ change is roughly symmetric around the equator, with slightly higher depletion in the North. Here, and in the figures that follow, time 0 is chosen at an arbitrary day after equilibrium is reached. The UVB ratio ranges from about 1.1 to just over 2.8. Note the large seasonal variation at high latitudes, where the change in sun angle is great.

Figure 4 presents the ratio of UVA irradiance comparing the SN case to the control. Note that the ratio values here are much smaller than for UVB. Also note that UVA shows both increase and decrease, depending on location and time of year. An increase is caused by less absorption by $O_3$, while a decrease is caused by more absorption by $NO_2$. While the change is quite small, and UVA is less damaging than UVB, this reduction may slightly offset an increase in damage from UVB.

Figure 5 presents the ratio of PAR. Values are larger everywhere in the SN case due to reduced absorption by $O_3$ The change here is larger than for UVA, but still quite small compared to UVB. An increase in PAR may be expected to lead to greater photosynthesis rates. Here the largest increase in PAR is at high latitudes, which could provide some benefit to plants that normally have short growing seasons with lower irradiance available. This again may offset the increase in damage from UVB.

The effects on organisms from changes in irradiance can be quite complicated, and in most cases is strongly wavelength dependent. Fortunately, we can simplify our view of the biological impacts by using weighting functions. In Figures 6-9, we show ratio plots for DNA damage (Setlow, 1974), erythema (Anders et al., 1995), skin cancer in humans (de Gruijl & Van der Leun, 1994), and cataracts in pigs (Oriowo et al., 2001). Note these figures use the same y-axis scale as that for UVB (Figure 3). Of these, the DNA damage ratios are highest; this is not surprising given that the Setlow function measures damage to "bare" DNA, while other effects are in organisms that have some shielding and/or repair mechanisms.

Effects on primary productivity may be even more relevant to ecosystem-wide effects. In Figures 10-12, we show ratio plots for photosynthesis inhibition in phytoplankton. Note these figures use the same y-axis scale, but that scale is smaller than in Figures 6-9. The largest increase is seen in Figure 10, for inhibition of carbon fixation in a natural Antarctic phytoplankton community, using the BWF from Boucher et al. (1994). Figures 11 and 12 show results using BWFs from Cullen et al. (1992), for inhibition of photosynthesis in the phytoplankton species *Phaeo-dactylum* ("phaeo") and *Prorocentrum micans* ("proro"), respectively.

For land plants, we find two widely different results, depending on weighting function chosen. Figure 13 shows results using a "generalized plant damage" BWF from Caldwell (1971). Figure 14 shows results using a BWF for damage to oat (*Avena sativa* L. cv. Otana) seedlings from Flint and Caldwell (2003). Note the y-axis scale on Figure 13 is much larger than that in Figure 14.

4. Discussion and Conclusions

Our goal in this work has been to investigate the potential biological impacts at Earth's surface of stratospheric $O_3$ depletion caused by nearby supernovae known to have occurred about 2.5 and 8 million years ago. We combined new and previously published atmospheric chemistry profiles with radiative transfer modeling to determine changes in surface-level UVB and biological responses. We find that UVB irradiance increased by a factor of 1.1 to 2.8, with large variation in latitude, and seasonally at the high latitude regions. Other wavelength bands of interest (UVA, PAR) are changed by relatively small amounts. Note that the estimate of UVB irradiance increase (1.5; see section 2) in Melott et al. (2017), computed using the RAF scaling method, fits in this range. We conclude that the RAF method is a reasonable approach for a first estimate of damage caused by $O_3$ depletion, which has the advantage of being quick and simple to apply, while full radiative transfer modeling can elucidate more nuanced details.

In our results, DNA damage shows an increase similar in magnitude to UVB. However, the DNA damage weighting function was determined under experimental conditions unlikely to exist in nature. We therefore examined other weighting functions, including erythema, skin cancer, and cataracts which are most relevant to larger organisms, possibly including early human ancestors (existing in East Africa around the 2.5 Ma SN events). We find increases ranging from a factor of about 1.1 to 2.0, again depending on latitude and season.

For marine phytoplankton (important organisms due to their role in global primary productivity) we find a much smaller impact (considering inhibition of photosynthesis), with variation between specific BWFs. Damage to land plants is widely different between the two BWFs considered here. The "generalized plant damage" BWF (Caldwell, 1971) shows more than 4 times the background damage (at 55° N), while the BWF for oat seedling damage (Flint & Caldwell, 2003) shows an increase of only about a factor of 1.15. Note that this latter value is similar in magnitude to that computed for two phytoplankton species (Figures 11 an 12).

An interesting feature of our results may be noted in comparing Figures 3, 4 and 5. Figure 3 shows a rather uniform cycle of increased UVB with seasons (due to changes in sun angle), most pronounced at higher latitudes. Figures 4 and 5 show the result of convolving that UVB irradiance with two biological weighting functions. Our expectation may be that UVB and (for instance) DNA damage changes would track exactly, but they clearly do not at the higher latitudes. This mis-match is caused by the fact that "UVB" is a broad-band integrated value, while DNA damage (as well as our other results) is a convolution of spectral irradiance and the (spectral) biological weighting function; therefore, differences in exactly which wavelengths are affected by changes in atmospheric chemistry impact the subsequent biological effects. We illustrate this in Figure 15, which shows the ratio of spectral irradiance comparing the SN case to a control run, at 55°N latitude, at day numbers 115 and 185. (Note that these day numbers are arbitrary and given simply to aid comparison, as discussed above.) Day number 115 corresponds to a time where UVB and DNA damage track together, while day number 185 corresponds to a time when UVB is still high but DNA damage has dropped. The explanation for this apparent discrepancy is that at day 185 there is less of an enhancement in the shorter wavelengths compared to day 115 and shorter wavelengths are much more effective at causing biological damage. Therefore, the result is less of an enhancement in damage at day 185 compared to day 115. This result motivates the use of full spectral modeling as opposed to simple broad-band scaling estimates, which may miss these features.

A similar effect can be seen in the plant damage results. First, the Caldwell (1971) "generalized plant damage" shows more than a factor of 4 increase over the control case, while UVB irradiance reaches a factor of not quite 3. It may seem that the increase in UVB-caused damage of any sort must be smaller than the increase in UVB itself. However, as Figure 15 again shows, the increase in irradiance is strongly wavelength-dependent. And, of course, so are BWFs. As may be seen in Figure 3 of Flint & Caldwell (2003), the generalized plant damage weighting function drops off very steeply at wavelengths above about 305 nm. It is therefore disproportionately affected by the shortest UVB wavelengths which are increased by as much as two orders of magnitude, while longer wavelengths are increased by much smaller factors.

This wavelength dependence explains the large difference in plant damage between the BWFs from Caldwell (1971) and Flint & Caldwell (2003). As noted above, the Caldwell (1971) BWF drops off steeply above about 305 nm and ends at about 315 nm. The Flint & Caldwell (2003) BWF is comparatively smaller in the 300-310 nm range and continues on from 310 to 370 nm. Recall from Figure 4 that the UVA (315-400 nm) irradiance actually *decreased* in the SN case, which therefore *reduces* the damage computed using the Flint & Caldwell (2003) BWF. These features again point out the importance of full, spectral radiative transfer modeling.

In order to address the question of how much of an impact on the biosphere our results represent, we need to put the increases in damage seen here in context. The best present-day analog for the scenario considered here is the anthropogenic ozone "hole" over Antarctica. During greatest extent, reductions of 50-60% are typical. While our SN case results have smaller depletion values, the duration and spatial extent are much larger, with near global coverage, sustained for centuries.

A number of studies have measured the impact on marine primary producers by depletion of $O_3$. We have surveyed the literature and here attempt to give a sense of scale for our results. For $O_3$ depletion of 50-60%, several studies found marine phytoplankton productivity around Antarctica decreased by few percent (Arrigo, 2003; Boucher & Prézelin, 1996; Neale et al., 1998; Smith et al., 1992). However, Smith et al. (1992) also point out that natural variability in productivity can be around 25%, due to changes in solar irradiance, water temperature, nutrient availability, etc. Our results (Figures 10-12) show increases of inhibition of phytoplankton productivity ranging from a few percent to 30-40%. We can conclude that this inhibition increase could have a noticeable, but unlikely catastrophic impact, just as the recent Antarctic changes have not had a catastrophic impact on that ecosystem (Karentz & Bosch, 2001). In particular, one might expect a change in relative species abundance in the ecosystem, as a result of the differential impact seen in our results (e.g. Figure 10 versus Figure 11).

Llabres & Agusti (2006) measured the 50% lethal doses from integrated UV radiation for two abundant phytoplankton species, *Procholorococcus* and *Synechococcus*. They found that for 7-8 hour exposures, irradiances of about 11 W m$^{-2}$ and 17 W m$^{-2}$ led to 50% mortality in these two species. These irradiance values are for total, integrated UV radiation, from 300-400 nm. The median irradiances we model are about 1 W m$^{-2}$ for UVB and 40 W m$^{-2}$ for UVA. Therefore, our scenario is likely to induce significant mortality in these two important species. Again, one might expect a change in relative species abundance in the ecosystem, with potentially significant reduction in abundance of these two species.

Several studies have examined effects of $O_3$ depletion on land plants, especially food crops. Searles et al. (2001) conducted a meta-analysis of 62 studies; they conclude that 20% $O_3$ depletion leads to 9-14% reduction in vegetative biomass. This level of $O_3$ depletion is seen over a wide range of latitude in our simulation results (Figure 2). Wu et al. (2009) found that a 30% increase in UVB radiation decreased leaf photosynthesis by 25-46% in several corn cultivars (but, had no significant effects in two cultivars). A similar or larger increase in UVB is found in our results (Figure 3), and so we may expect an impact on terrestrial photosynthesis as well, with again the caveat that there is variation between plant species. Tevini (1993) reports that for 12% $O_3$ depletion (corresponding to a 25% UVB increase) there is a continuous reduction in growth rates of sunflower plants, and a significant reduction in biomass for one third of rice cultivars under 20% $O_3$ depletion. Bornman & Teramura (1993) report that $O_3$ depletion ranging from 16-32% can cause large reduction in yields for plants such as squash, tomato, mustard and black-eyed pea, but that final yield of soybeans and corn are not impacted. Conversely, some soybean and wheat cultivars actually increased their yield under simulated 25% $O_3$ depletion. As with marine ecosystems, it appears the most likely result of our modeled $O_3$ changes would be modification of species abundances, with some species more negatively impacted than others.

We also emphasize that the increase in UV modeled here lasts for centuries or longer and is widespread across the globe. The greatest increases are seen at the highest latitudes, which cover a smaller area and tend to host lower diversity of organisms; however, these high latitudes are still important areas of seasonal productivity, relied upon by many migrant species such as whales, which will also be exposed directly to the higher UV irradiance with negative consequences (Martinez-Levasseur et al., 2011).

Overall, we can conclude that changes in UV irradiance at Earth's surface following the supernovae known to have occurred around 2.5 and 8 million years ago most likely had the effect of changing ecosystem balances, significantly impacting some species with little negative (and possibly even positive) impacts on others. This conclusion fits well with the observations that no major mass extinctions occurred around these time periods, but that there was elevated extinction, turnover in species, and changes in vegetation cover. For instance, Behrensmeyer et al. (1997) report that 58-77% of mammal species were replaced between 3.0 and 1.8 million years ago, with the most significant period of faunal change between 2.5 and 1.8 Ma. In addition, there was a marked increase in grasslands at the expense of forests (Pennington & Hughes, 2014).

Our results may motivate ecosystem-level modeling that takes the pressures found in our results into account. It should also be noted that in this work we do not consider other potential biological impacts, most notably increased radiation on the ground due to cosmic ray muons. Discussion of the potential for increased (internal) cancer rates due to this effect may be found in Thomas et al. (2016) and Melott et al. (2017).

Acknowledgments

The author gratefully acknowledges support from the NASA Exobiology program, grant #NNX14AK22G. Thanks to Mikhail Medvedev for pointing out that Moscow and St. Petersburg, Russia, are at 55° and 60° North latitude, so that high latitude effects should not be ignored! Computational time was provided by the High Performance Computing Environment (HiPACE) at Washburn University; thanks to Steve Black for assistance with computing resources.

Author Disclosure Statement

No competing financial interests exist.

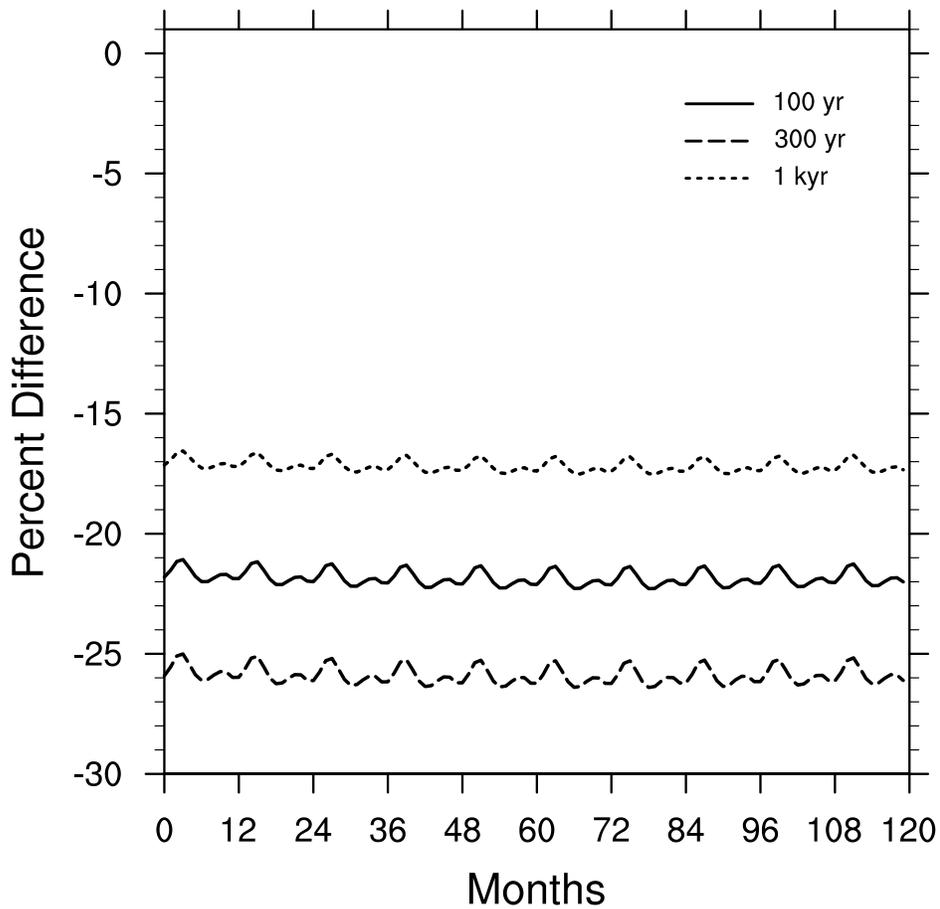

Figure 1

Globally averaged change in O₃ column density, as a percent difference (comparing model run with a constant SN CR input to a control run), showing the steady state for each era (100 yr, 300 yr, 1 kyr). Time 0 indicates an arbitrary reference time after steady state is achieved.

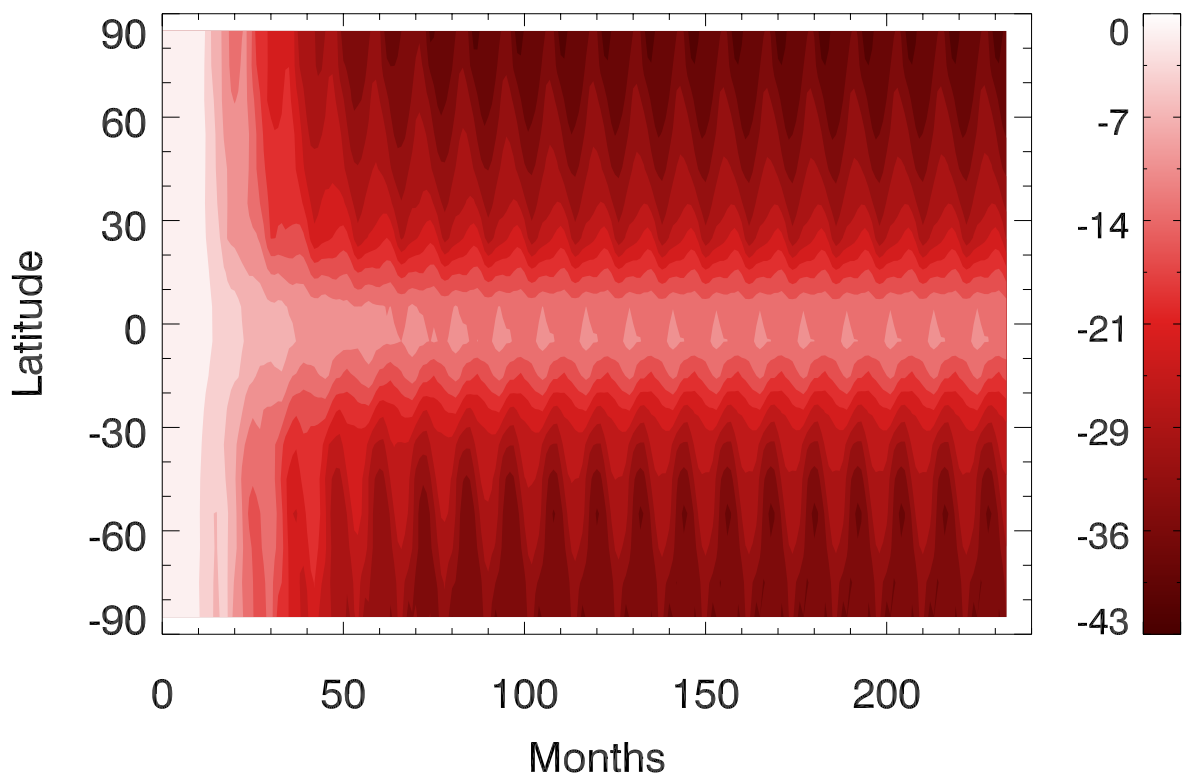

Figure 2

Percent difference in O₃ column density as a function of latitude and time for the 300 yr SN case (versus control). The time axis values are measured from the beginning of the simulation run, with SN CR ionization input beginning one year after the start of the run.

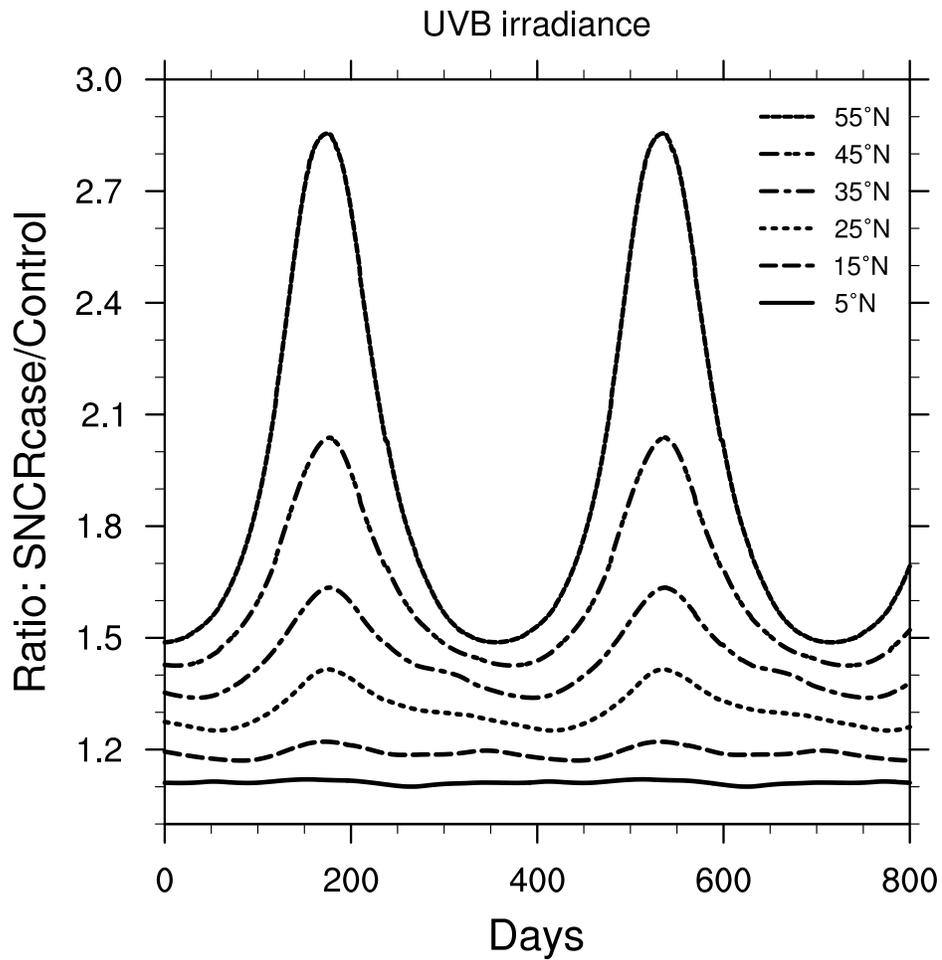

Figure 3

Ratio of surface-level UVB irradiance in the 300 yr SN case compared to control run, as a function of time, at latitudes between 5° and 55° North. Time 0 indicates an arbitrary reference time after steady state is achieved.

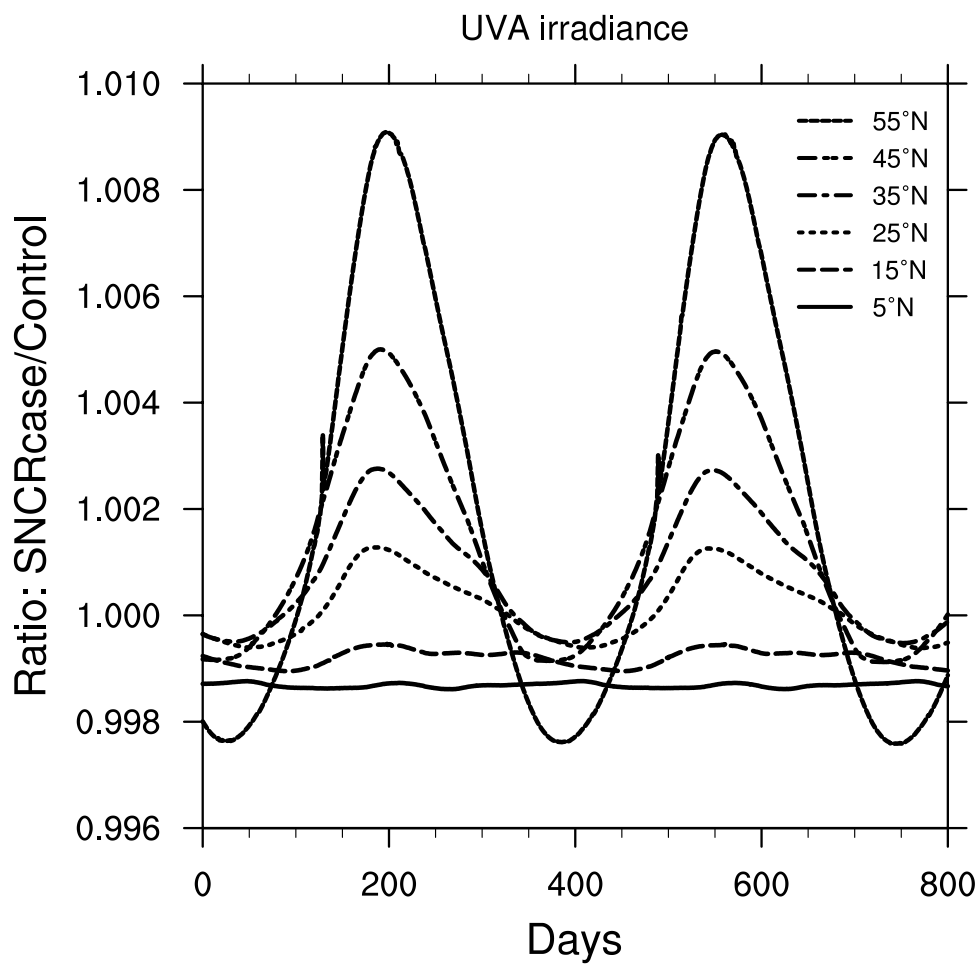

Figure 4

Ratio of surface-level UVA irradiance in the 300 yr SN case compared to control run, as a function of time, at latitudes between 5° and 55° North.  Time 0 indicates an arbitrary reference time after steady state is achieved, and is the same start time as in Figure 3.

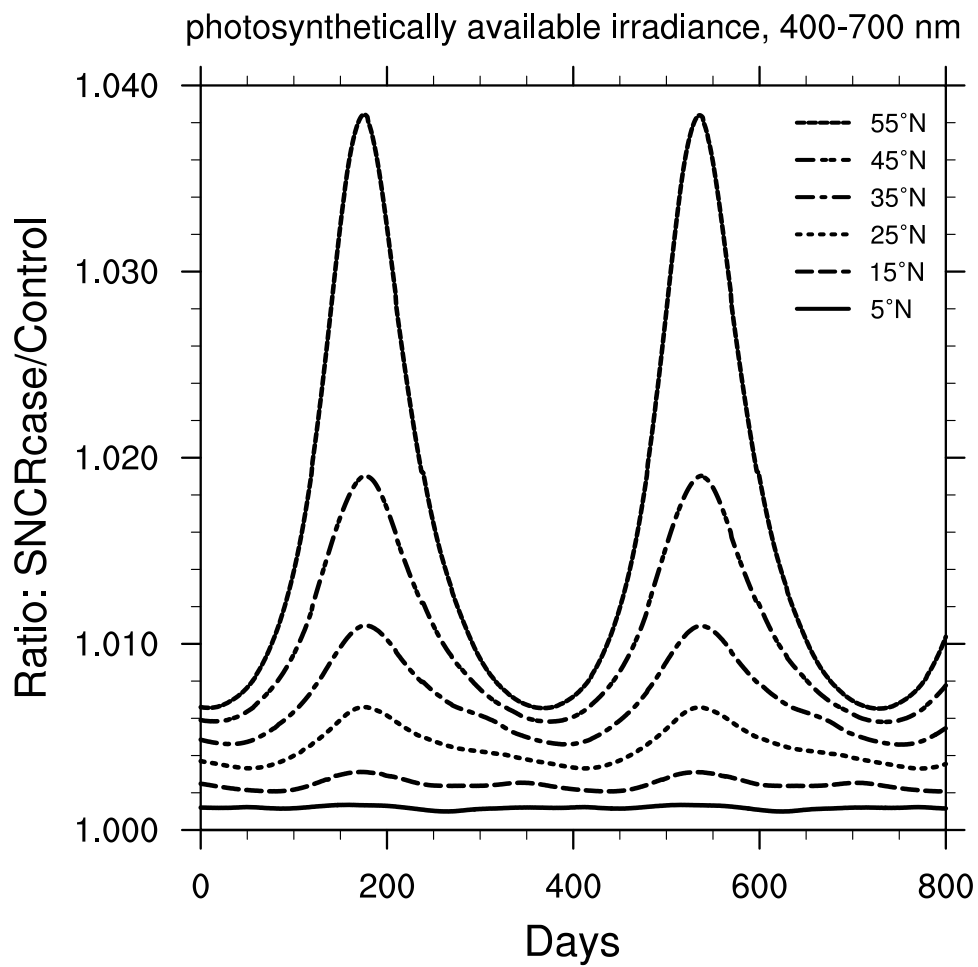

Figure 5

Ratio of surface-level PAR irradiance in the 300 yr SN case compared to control run, as a function of time, at latitudes between 5° and 55° North. Time 0 indicates an arbitrary reference time after steady state is achieved, and is the same start time as in Figure 3.

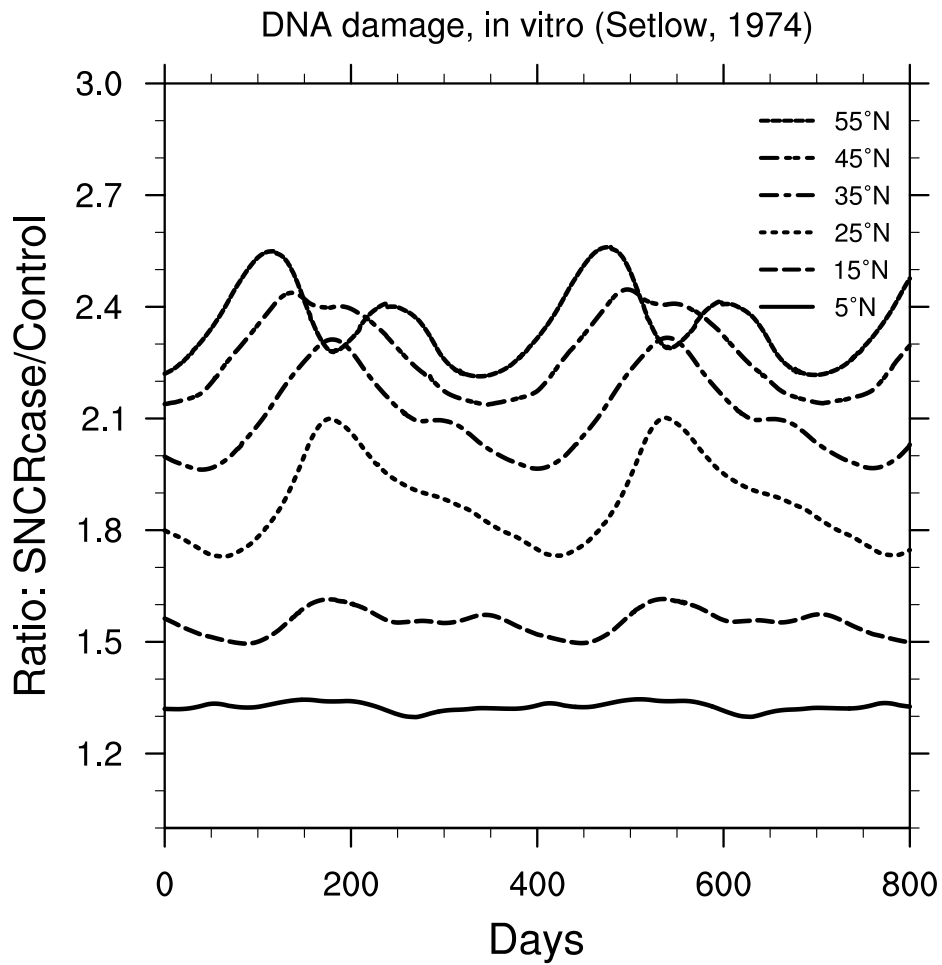

**Figure 6**

Ratio of surface-level DNA damage (Setlow, 1974) in the 300 yr SN case compared to control run, as a function of time, at latitudes between 5° and 55° North. Time 0 indicates an arbitrary reference time after steady state is achieved, and is the same start time as in Figure 3.

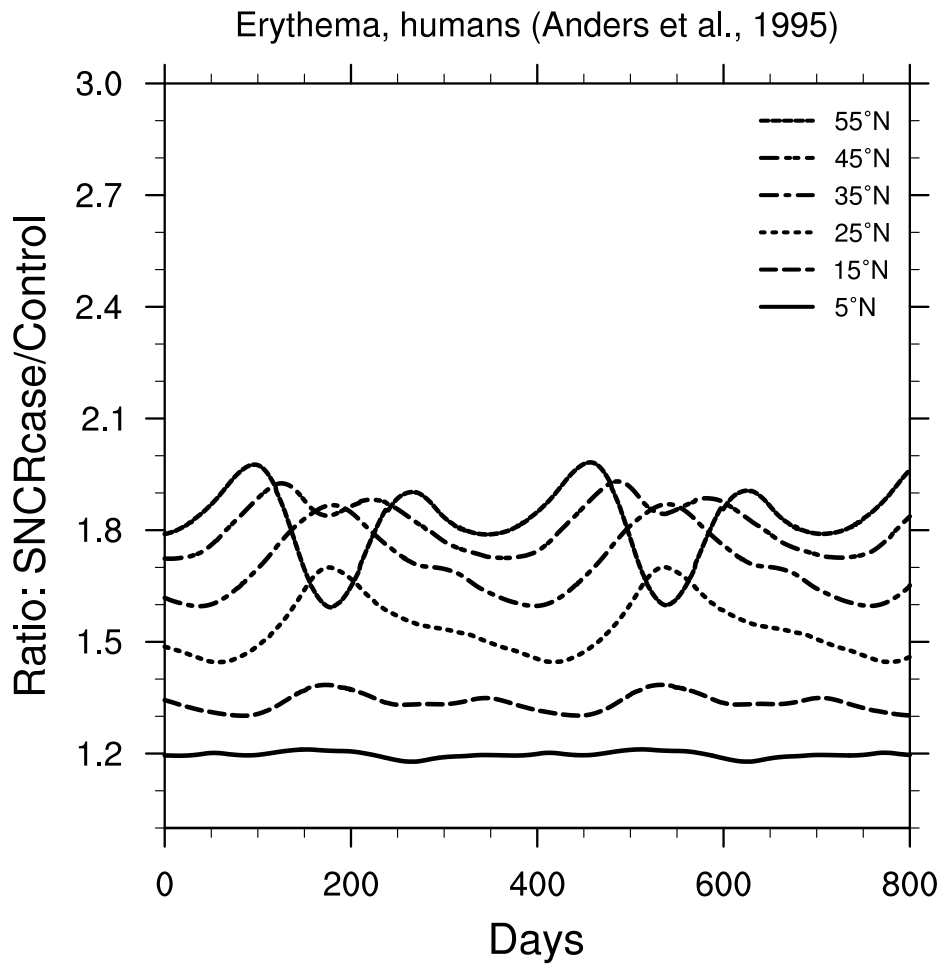

Figure 7

Ratio of surface-level erythema in humans (Anders et al., 1995), in the 300 yr SN case compared to control run, as a function of time, at latitudes between 5° and 55° North. Time 0 indicates an arbitrary reference time after steady state is achieved, and is the same start time as in Figure 3.

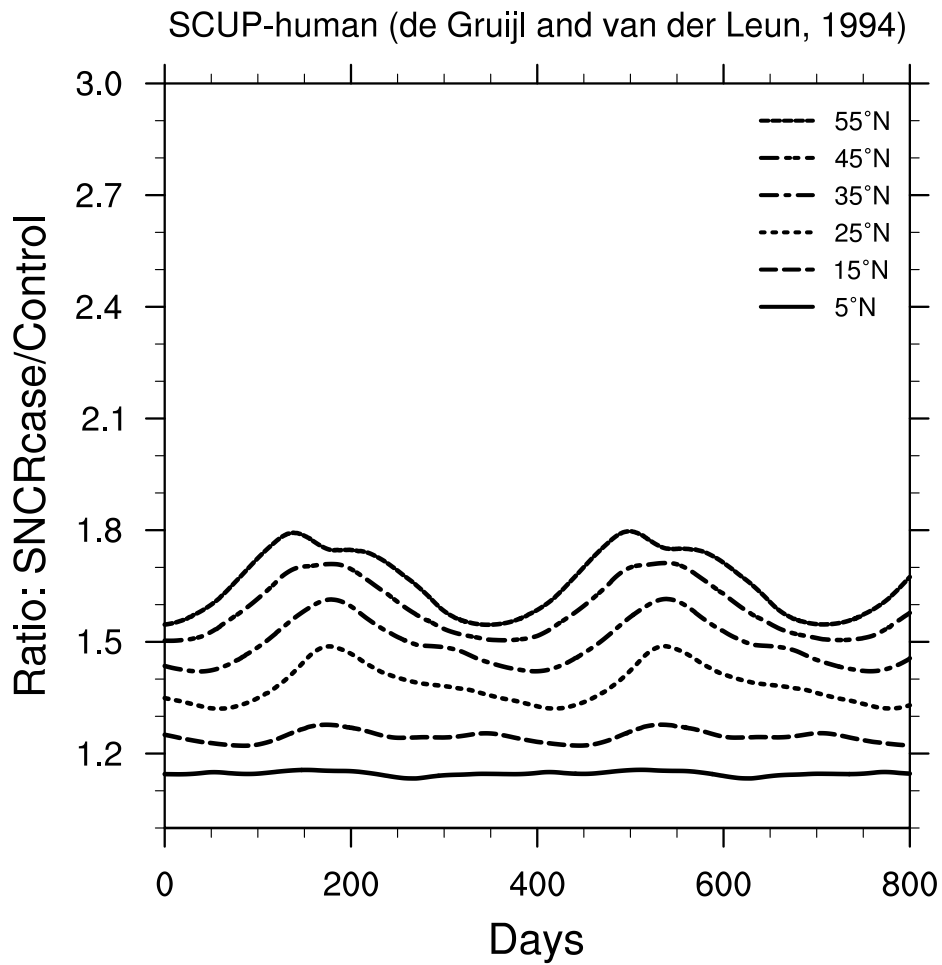

Figure 8

Ratio of surface-level skin cancer in humans (de Gruijl & Van der Leun, 1994), in the 300 yr SN case compared to control run, as a function of time, at latitudes between 5° and 55° North. (Note that "SCUP" stands for "Skin Cancer Utrecht-Philadelphia.") Time 0 indicates an arbitrary reference time after steady state is achieved, and is the same start time as in Figure 3.

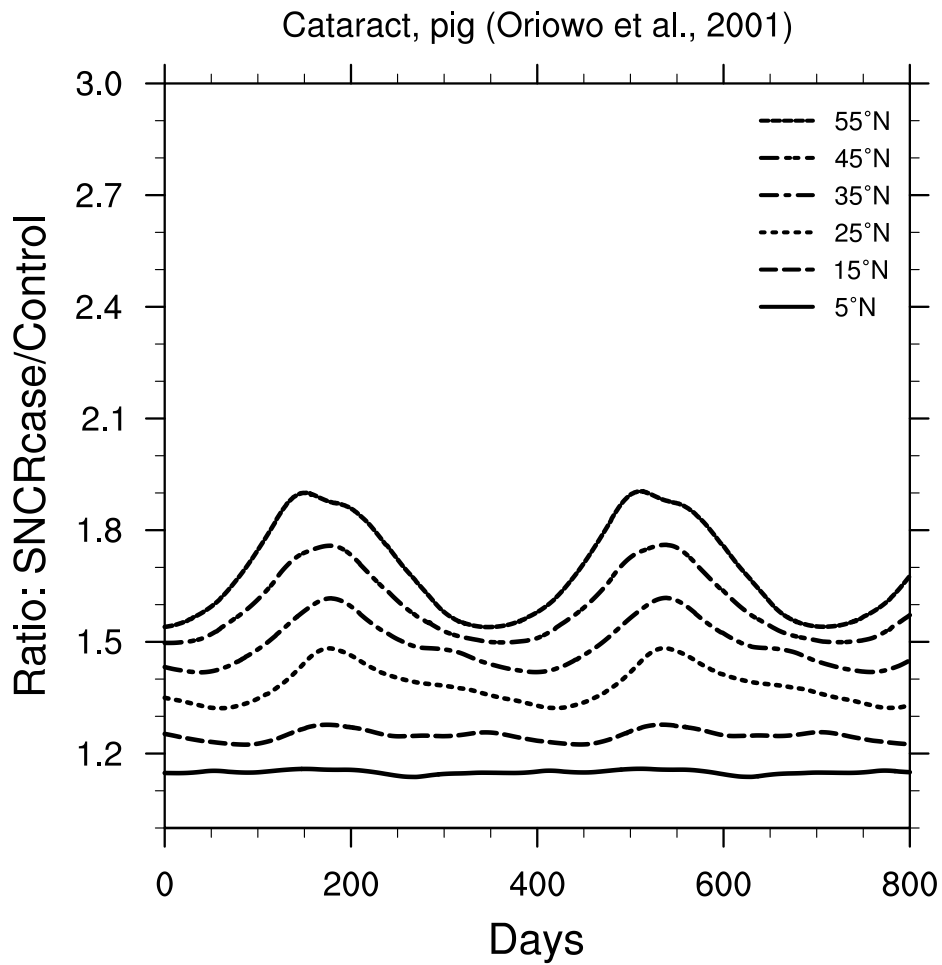

Figure 9

Ratio of surface-level formation of cataracts in pig eye lenses (Oriowo et al., 2001), in the 300 yr SN case compared to control run, as a function of time, at latitudes between 5° and 55° North. Time 0 indicates an arbitrary reference time after steady state is achieved, and is the same start time as in Figure 3.

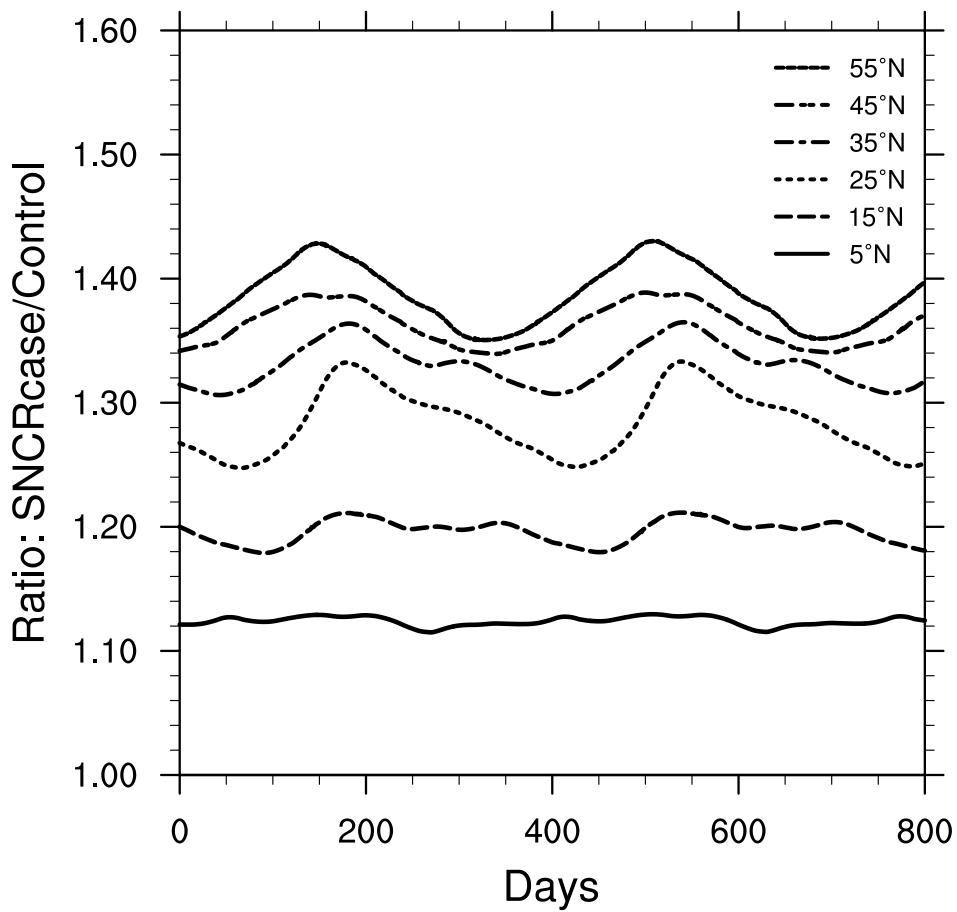

**Phytoplankton inhibition of primary production (Boucher et al., 1994)**

Figure 10

Ratio of surface-level inhibition of carbon fixation in a natural Antarctic phytoplankton community (Boucher et al., 1994), in the 300 yr SN case compared to control run, as a function of time, at latitudes between 5° and 55° North.  Time 0 indicates an arbitrary reference time after steady state is achieved, and is the same start time as in Figure 3.

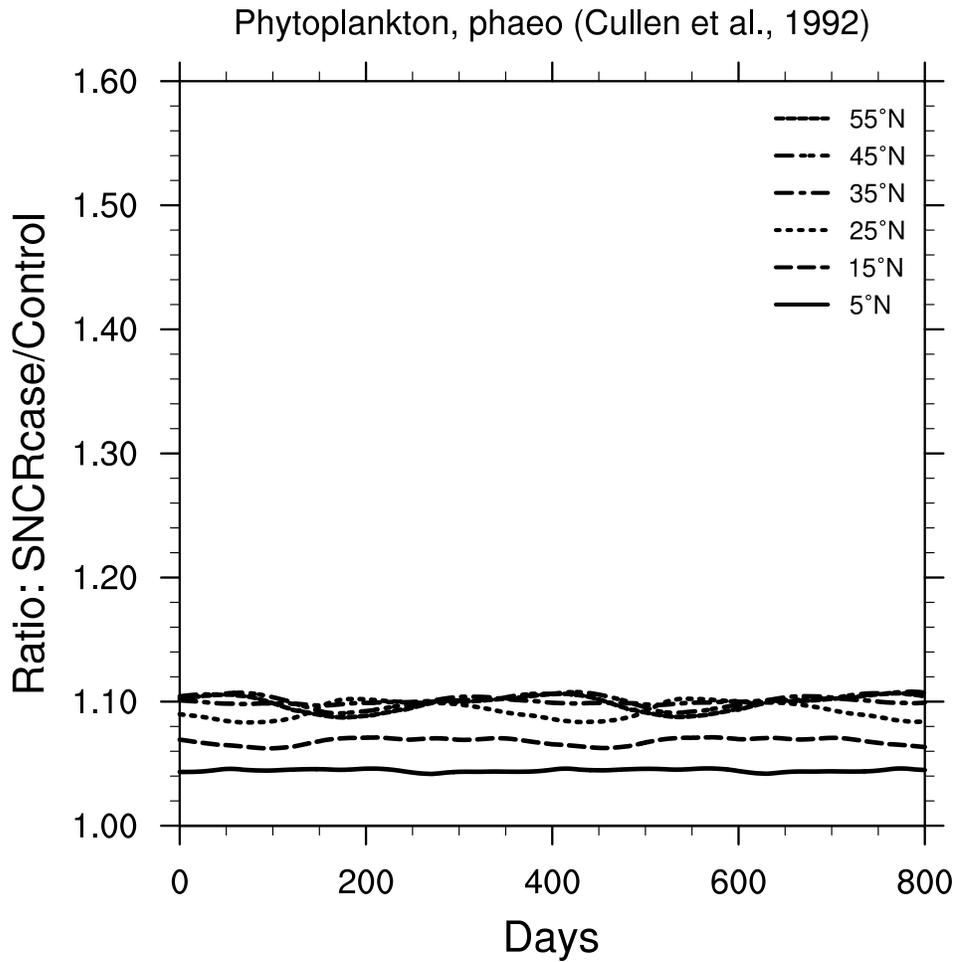

Figure 11

Ratio of surface-level inhibition of photosynthesis in the phytoplankton species *Phaeo-dactylum* (Cullen et al., 1992), in the 300 yr SN case compared to control run, as a function of time, at latitudes between 5° and 55° North.  Time 0 indicates an arbitrary reference time after steady state is achieved, and is the same start time as in Figure 3.

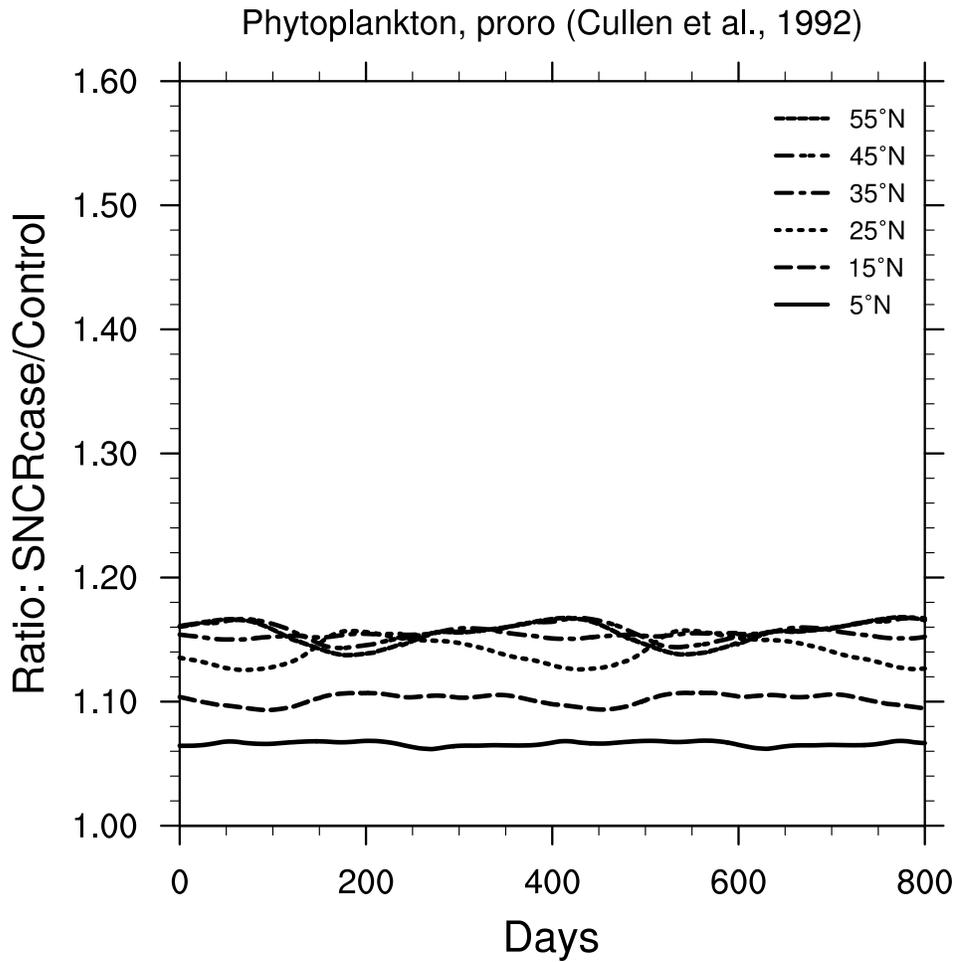

Figure 12

Ratio of surface-level inhibition of photosynthesis in the phytoplankton species *Prorocentrum micans* (Cullen et al., 1992), in the 300 yr SN case compared to control run, as a function of time, at latitudes between 5° and 55° North.  Time 0 indicates an arbitrary reference time after steady state is achieved, and is the same start time as in Figure 3.

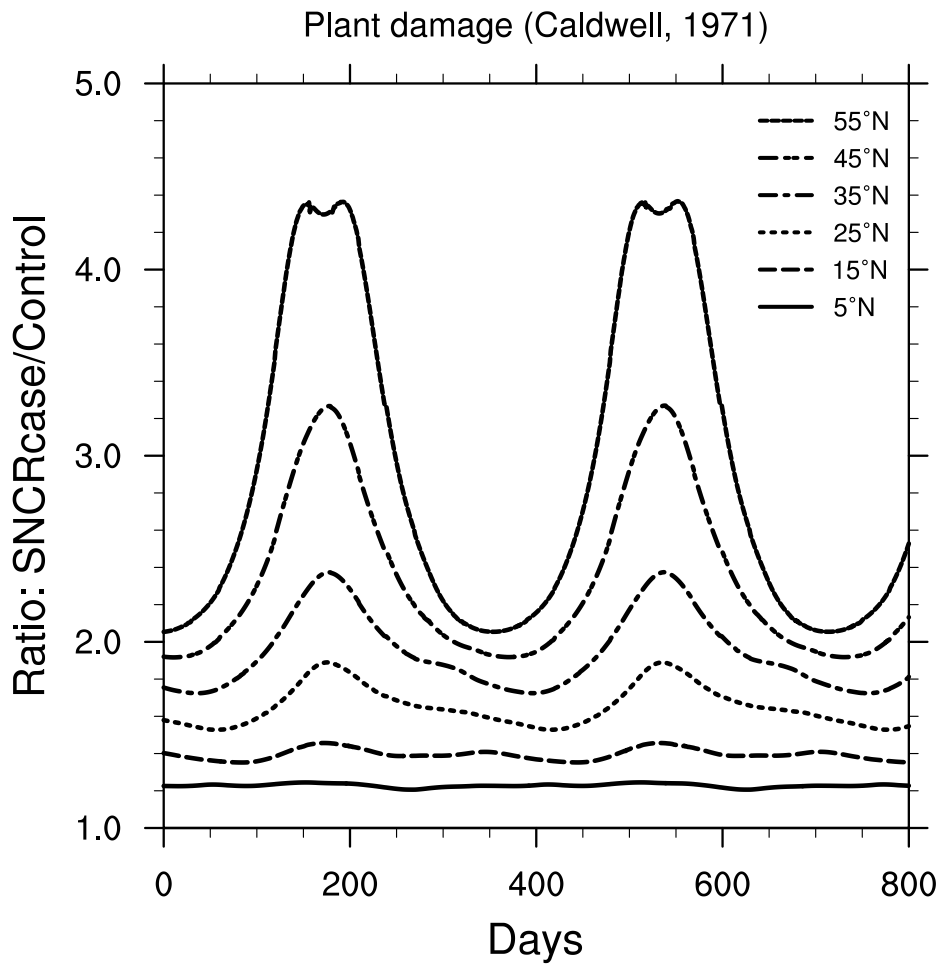

Figure 13

Ratio of surface-level generalized plant damage (Caldwell, 1971), in the 300 yr SN case compared to control run, as a function of time, at latitudes between 5° and 55° North. Time 0 indicates an arbitrary reference time after steady state is achieved, and is the same start time as in Figure 3.

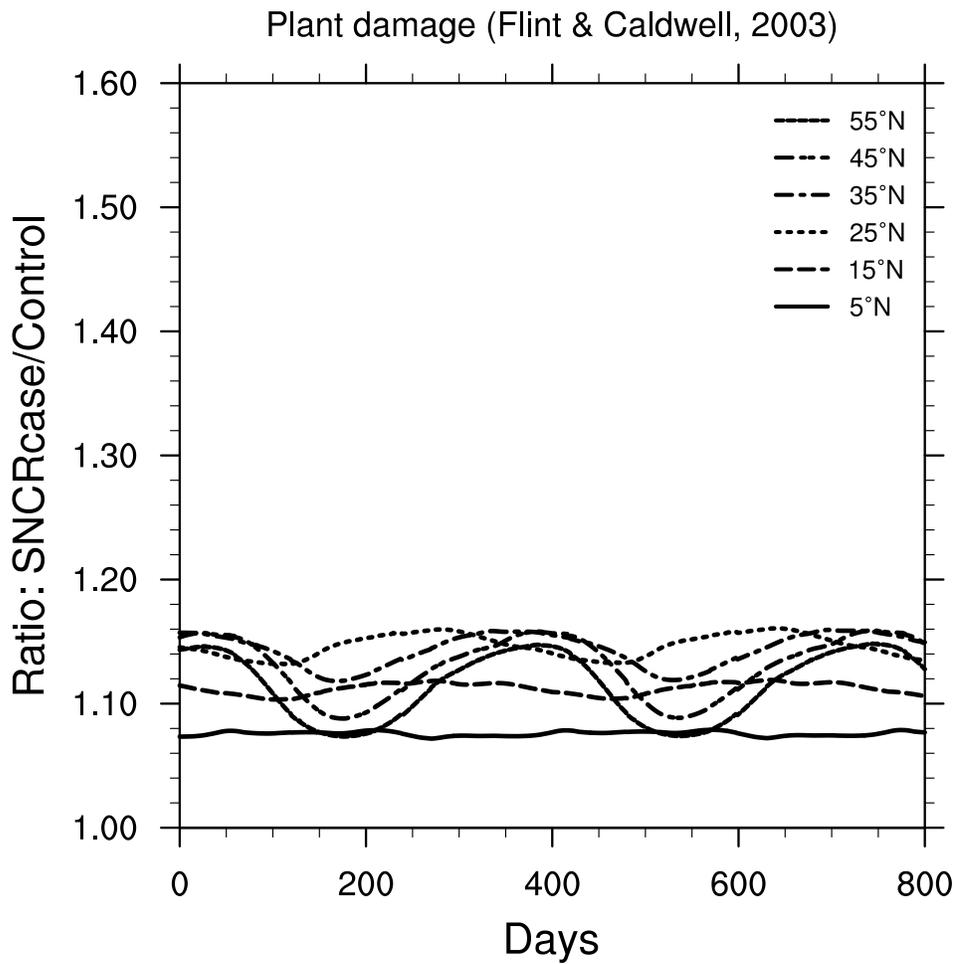

**Figure 14**

Ratio of surface-level damage to oat (*Avena sativa* L. cv. Otana) seedlings (Flint and Caldwell, 2003), in the 300 yr SN case compared to control run, as a function of time, at latitudes between 5° and 55° North. Time 0 indicates an arbitrary reference time after steady state is achieved, and is the same start time as in Figure 3.

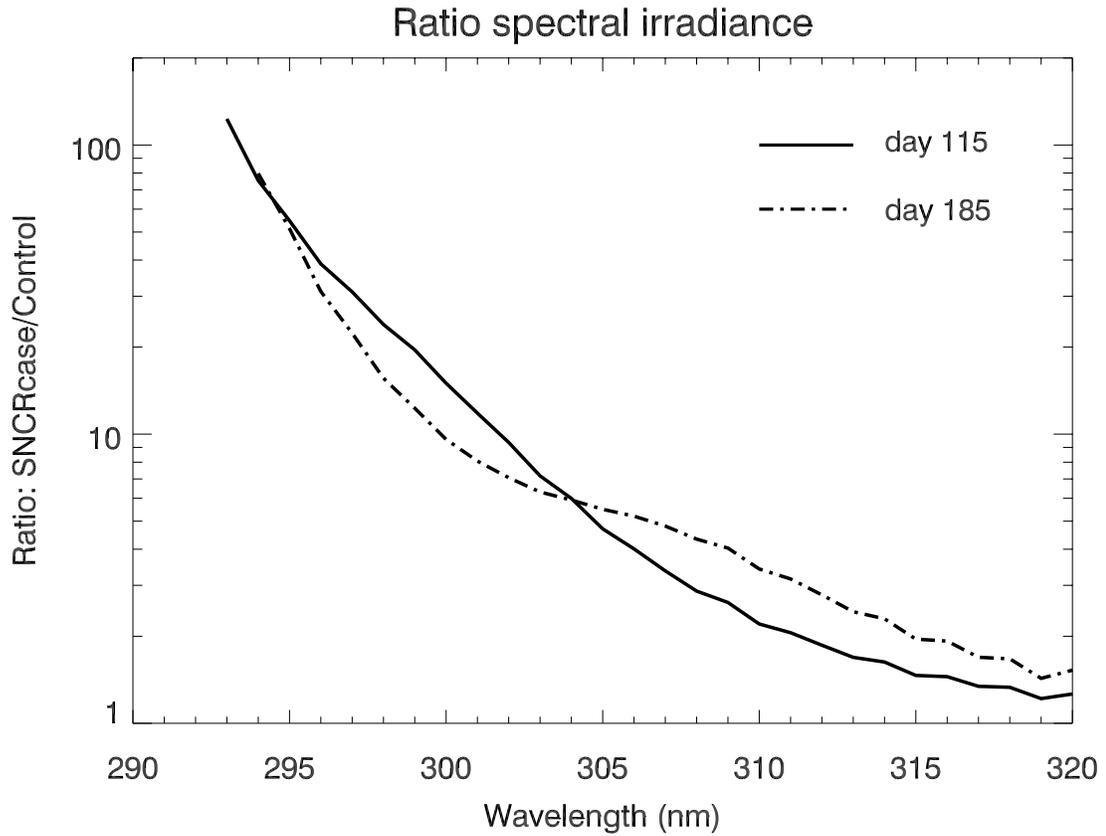

Figure 15

Ratio of spectral irradiance in the UVB band comparing the SN case to a control run, at 55°N latitude, at day numbers 115 and 185. (Day numbers are measured from time 0, an arbitrary reference time after steady state is achieved, which is the same start time as in Figure 3.)